\documentclass[preprint]{aastex}

\usepackage{graphicx}
\usepackage[usenames,dvipsnames]{color}
\usepackage[normalem]{ulem}

\begin{document}
\title{Is FS Tau B Driving an Asymmetric Jet?\altaffilmark{*}}
\author{Chun-Fan Liu\altaffilmark{1,2,3}, Hsien Shang\altaffilmark{2,3}, 
        Tae-Soo Pyo\altaffilmark{4}, Michihiro Takami\altaffilmark{2}, 
        Frederick M. Walter\altaffilmark{5},
        Chi-Hung Yan\altaffilmark{2,6}, Shiang-Yu Wang\altaffilmark{2}, 
        Nagayoshi Ohashi\altaffilmark{2,4}, \& Masahiko Hayashi\altaffilmark{7}}
\altaffiltext{*}{Based on data collected at Subaru Telescope, 
                 which is operated by the National Astronomical Observatory of Japan.}
\altaffiltext{1}{Graduate Institute of Astrophysics, National Taiwan University, 
                 No. 1, Sec. 4, Roosevelt Road, Taipei 10617, Taiwan}
\altaffiltext{2}{Institute of Astronomy and Astrophysics, Academia Sinica (ASIAA),
                 P.O. Box 23-131, Taipei 10641, Taiwan}
\altaffiltext{3}{Theoretical Institute of Advanced Research in Astrophysics (TIARA)}
\altaffiltext{4}{Subaru Telescope, National Astronomical Observatory of Japan,
                 650 North A'ohoku Place, Hilo, HI 96720, USA}
\altaffiltext{5}{Department of Physics and Astronomy, Stony Brook University, 
                 Stony Brook, NY 11794-3800, USA}
\altaffiltext{6}{Department of Earth Sciences, National Taiwan Normal University, 
                 Taipei 11677, Taiwan}
\altaffiltext{7}{Department of Astronomy, School of Science, University of Tokyo,
                 Tokyo 113-0033, Japan}
\slugcomment{Accepted for publication in ApJ}

\begin{abstract}
 FS Tau B is one of the few T Tauri stars that possess a jet 
 and a counterjet as well as an optically-visible cavity wall.
 We obtained images and spectra of its jet-cavity system in the near-infrared $H$ and $K$ 
 bands using Subaru/IRCS and detected the jet and the counterjet 
 in the [Fe {\sc ii}] 1.644 \micron\ line for the first time. 
 Within the inner 2\arcsec\ the blueshifted jet is brighter, whereas beyond $\sim 5\arcsec$
 the redshifted counterjet dominates the [Fe {\sc ii}] emission. 
 The innermost blueshifted knot is spectrally resolved to have a large 
 line width of $\sim 110$ km\,s$^{-1}$, while the innermost redshifted knot appears 
 spectrally unresolved. The velocity ratio of the jet to the counterjet is $\sim 1.34$, 
 which suggests that FS Tau B is driving an asymmetric jet, similar to those 
 found in several T Tauri Stars. 
 Combining with optical observations in the literature, we showed that the
 blueshifted jet has lower density and higher excitation than the redshifted counterjet.
 We suggest that the asymmetry in brightness and velocity is the manifestation of
 a bipolar outflow driving at different mass-loss rates, while maintaining balance of
 linear momentum. A full explanation to the asymmetry in the FS Tau B system
 awaits detail modeling and further investigation of the kinematic structure of the 
 wind-associated cavity walls.
\end{abstract}


\section{Introduction}
 Recent observational progress in the study of jets and outflows from low-mass young 
 stellar objects (YSOs) has focused on revealing how outflows are driven from close to 
 the star and its inner disk. 
 In the stage of T Tauri Stars (TTSs), the YSOs have exposed their central stars and 
 the close vicinity in the optical wavelengths. The TTSs are known to drive jets 
 in optical forbidden emission lines.
 High spatial resolution spectra obtained from STIS on the {\it Hubble Space Telescope (HST)}
 have provided detailed spatial and velocity distributions of emission lines along and across the jets.
 Information on physical conditions of several jets within 
 hundreds of AU from the driving region has been extracted from atomic lines such as [O {\sc i}], 
 [S {\sc ii}], and [N {\sc ii}] (Bacciotti et al. 2000; Woitas et al. 2002b;
 Hartigan \& Morse 2007; Coffey et al. 2008; Melnikov et al. 2009).

 The [Fe {\sc ii}] 1.644 \micron\ emission line provides diagnostics probing deeper into the 
 driving source. [Fe {\sc ii}] traces shock as do optical forbidden emission lines but suffers
 less from circumstellar extinction (Hollenbach \& McKee\ 1989). 
 It has been shown from {\it HST}/NICMOS images that [Fe {\sc ii}] traces shocked Herbig-Haro
 objects close to the more embedded Class I sources (see discussion in Reipurth et al.\ 2000).
 Long-slit near-infrared spectroscopy has shown that kinematic properties of the [Fe {\sc ii}]
 line is similar to those seen in the optical [S {\sc ii}] line and are important to both Class  
 I and II (TTSs) sources (Davis et al.\ 2003).
 High-dispersion echelle spectrographs aided with adaptive optics provide spatially resolved 
 [Fe {\sc ii}] spectra down to scale of 15 AU from the source. The properties of the line profiles 
 along the jet enable opportunities for confrontation with models
 (Pyo et al.\ 2003, 2006; Hartigan \& Hillenbrand\ 2009).

 Attention has been drawn to the studies of differences in physical properties between the 
 jet and the counterjet of several TTS systems in which both sides of the bipolar jet were detected.
 Hirth et al. (1994) conducted a long-slit optical spectroscopic survey of TTS jets and found that 
 in 8 out of 15 sources the velocity ratios between the faster side and the slower side of the jet 
 range from 1.4 to 2.6 while in other cases the velocity differences are less significant.
 Besides asymmetric velocity, forbidden line ratios that trace electron density and 
 temperature also differ between the two sides of jet lobes (Hirth et al. 1997). 
 The jet from RW Aur A is one of the examples that have been extensively studied.
 It is fainter but faster on the blueshifted side in counterpoint to the 
 brighter redshifted side (Woitas et al. 2002b; Melnikov et al. 2009). 
 {\it HST} and ground-based telescopes aided with adaptive optics spatially resolved 
 the jet and showed that the non-unity velocity ratio between the counterjet and the jet 
 can be traced in to 0\farcs2 from the driving source in both the optical and 
 near-infrared spectra (Woitas et al.\ 2002b; Melnikov et al.\ 2009; Pyo et al.\ 2006;
 Hartigan \& Hillenbrand\ 2009). Line ratio diagnostics show that physical 
 conditions inside the jet and the counterjet differ greatly, and the ionization fraction 
 differs by at least a factor of two (Melnikov et al. 2009; Liu \& Shang 2012). 
 Understanding the origin of asymmetry close to the driving source would illuminate how mass 
 is loaded onto the outflow driven by a magnetocentrifugal wind.

 FS Tau B (Haro 6-5 B) offers another opportunity to examine the asymmetries in an 
 outflow system. It drives a bipolar jet at a position angle of $\sim 54\degr$, 
 in which the most prominent knot, known as HH 157, lies $\sim 30\arcsec$ from the 
 driving source, FS Tau B (Mundt et al. 1984; Eisl\"offel \& Mundt 1998). 
 The asymmetries between the jet and counterjet can be seen in various aspects.
 Ground-based imaging showed that the jet appears broader and extended longer than
 the more compact and straight counterjet (Mundt et al. 1991; Eisl\"offel \& Mundt 1998). 
 HST/WFPC2 broad-band imaging showed that the jet appears somewhat meandering and more diffuse than
 the counterjet (Krist et al. 1998). From optical narrow-band imaging, the jet appears
 more prominent in H$\alpha$ than in [S {\sc ii}], while the opposite is true in the 
 counterjet. This phenomenon can be seen from both ground-based imaging (Eisl\"offel \& Mundt 1998)
 and HST/WFPC2 images (Woitas et al. 2002a). Optical long-slit spectra show an asymmetry 
 in velocity (Mundt et al. 1987), with a blueshifted jet $\sim 20\%$ faster than the 
 counterjet if systemic velocity is considered (Eisl\"offel \& Mundt 1998).
 Although the velocity ratio of 1.2 leads Hirth et al. (1994) to categorize FS Tau B into the rather
 symmetric half of the samples, the difference observed in the narrow-band imaging suggests different
 physical conditions prevail on the two sides of the jet lobes. We investigate this aspect of 
 asymmetry in the near-infrared wavelengths using the [Fe {\sc ii}] 1.644 \micron\ line.
 An ovoid structure surrounding the northeastern jet shows polarization pattern 
 centered at FS Tau B (Gledhill \& Scarrott 1989) which may either be the outline of a 
 light cone (Krist et al. 1998) or the cavity wall carved by the FS Tau B outflow 
 (Eisl\"offel \& Mundt 1998). Deep imaging by Eisl\"offel \& Mundt (1998) and Krist et al. 
 (1998) revealed its much fainter southwestern counterpart. The asymmetry in the jet and in 
 the ovoid structure related to the outflow cavity wall makes the
 FS Tau B system a testbed for studying the asymmetry in outflows from young stellar systems.

 Although the asymmetry between the jet and the counterjet has been well documented,
 the origin of asymmetry was not fully understood. Mundt et al. (1987) 
 were inclined to attribute the asymmetry to different density distributions in the 
 surrounding molecular clouds on the two sides of the jet. Considering the difference 
 in initial opening of the jet and the counterjet, Mundt et al. (1991) acknowledged 
 that the difference cannot be explained by differential extinction on the two sides.
 They concluded that apart from any intrinsic asymmetry, the circumstellar environment 
 plays a fundamental role in shaping the outflow through interaction. The conclusion was 
 based on the observation that densest gas is often offset from the stellar core
 in the dark cloud, yet how the dense gas is distributed around FS Tau B is still not clear.
 To explain the difference in the brightness of the ovoid structures, Krist et al. (1998) 
 proposed that a dark cloud within the northeastern lobe provides denser material for 
 scattering. Overall, interpretations in literature have mainly focused on possible 
 environmental differences rather than on whether FS Tau B may be driving an intrinsically 
 asymmetric jet and outflow. Using near-infrared imaging and spectroscopy in [Fe {\sc ii}], 
 we studied the properties of the jet close to the driving source. By comparing 
 the near-infrared and optical results, we show that those asymmetric properties of 
 the FS Tau B system may be attributed to an asymmetric outflow driven from the source.
 
 The $X$-wind model provides a good framework for understanding the morphology 
 and kinematic features for jets and outflows from YSOs (Shu et al. 1994a, b; 
 Shang et al. 1998, 2002). The model describes a wide-angle magnetocentrifugal wind 
 that is driven from the inner edge of a circumstellar disk. 
 The density in the wind is magnetically collimated towards the axis and
 collisionally excites optical lines such as [O {\sc i}] and [S {\sc ii}] that appear
 as the high-speed jet revealed in TTS phase (Shang et al. 1998).
 The wide-angle nature of the wind is manifested by a very broad line shape excited
 close to the driving source (Shang et al. 1998, 2010). To compare the observation 
 with the model, obtaining spatially resolved spectra close to the source is crucial.

 In this paper, we report results from [Fe {\sc ii}] 1.644 \micron\ 
 spectroimaging of FS Tau B with the Subaru Telescope.
 In \S 2, we summarize the observations, and in \S 3, we show results of the spatially and 
 spectrally resolved [Fe {\sc ii}] jet.
 We discuss asymmetric properties of the jet and compare our 
 near-infrared results with those obtained in optical wavelengths from literature in \S 4.
 We summarize our findings in \S 5.

\section{Observations and Data Reduction}

\subsection{Subaru IRCS Imaging in $H$ and $K$ Bands}
 Images of FS Tau A/B were obtained on January 3, 2007, with the Infrared Camera and Spectrograph 
 (IRCS; Kobayashi et al. 2000; Tokunaga et al. 1998) on the Subaru Telescope. 
 Narrow-band filters were used to isolate [Fe {\sc ii}] (with central wavelength 
 $\lambda_{\rm c} = 1.644$ \micron\ and bandwidth $\Delta\lambda = 0.026$ \micron) 
 and H$_2$ ($\lambda_{\rm c} = 2.122$ \micron; $\Delta\lambda = 0.032$ \micron); 
 broad-band $H$ ($\lambda_{\rm c} = 1.63$  \micron; $\Delta\lambda = 0.30$  \micron) and 
 $K^\prime$ ($\lambda_{\rm c} = 2.12$  \micron; $\Delta\lambda = 0.35$  \micron) filters 
 were used to measure the continuum fluxes. The 52.42 mas/pixel scale yields a 
 53\farcs68 square field of view that covers the driving 
 region of HH 157. The telescope was nodded between the source and the sky 3\arcmin\ 
 southeast from the source. At each nodding position, the pointing was dithered in
 a five-point square dither pattern (S5) with dithering distance 5\farcs6. 
 The seeing on the summit of Mauna Kea was between 0\farcs8 and 1\farcs1. 
 The total on-source integration times for $H$ and $K^\prime$ filters were 90 and 162 s, 
 and 900 and 1980 s, respectively, for the [Fe {\sc ii}] and H$_2$ filters.

 We reduced the data with standard procedures, including flatfielding, bad pixel correction, 
 sky subtraction, ghost identification and replacement, dithered frame shifting, and image
 stacking. Latency and ghosts caused by the bright FS Tau A required iterative sky subtraction.
 Ghost pixels were identified on both sky and source frames and replaced with unaffected 
 patches by inspection. The ghost-free sky frames were then combined to obtain the final 
 sky frame, which was subtracted from each ghost-free on-source frame. 
 The reduced on-source frames were shifted and coadded to form the final images.

\subsection{IRCS Echelle Spectra with [Fe {\sc ii}] 1.644 \micron\ Line}
 IRCS echelle spectra of FS Tau B in the $H$-band were taken on November 27, 2002.
 The 0\farcs618 wide $\times$ 5\farcs79 long slit was placed along the jet 
 at the position angle of 55\degr. The velocity resolution is 60 km\,s$^{-1}$ 
 ($R \sim 5000$). The plate scales for spatial and dispersion axes are 0\farcs06 and 
 0.4044 \AA, respectively. The total on-source exposure time was 1200 s. The sky 
 spectra were obtained at positions $\sim15\arcsec$ northwest and southwest from 
 the source position. The seeing at $K$-band was 0\farcs55.

 The spectra were reduced using standard {\it IRAF}\footnote{IRAF is distributed by 
 the National Optical Astronomy Observatory, which is operated by the Association of 
 Universities for Research in Astronomy (AURA) under cooperative agreement with the 
 National Science Foundation.} packages for echelle spectra reduction, following the 
 procedure in Pyo et al. (2003). HR 1570 (A0V, $H=4.517$) was used as telluric and 
 flux calibration standard. The continuum 
 in the 34$^{\rm th}$ order of the echelle (wavelengths 1.635 to 1.675 \micron)
 was fit with a 5$^{\rm th}$ order Chebyshev polynomial 
 and subtracted using the {\tt BACKGROUND} procedure.

\section{Results}

\subsection{Collimated [Fe {\sc ii}] Counterjet from FS Tau B}

%%% Figure 1 %%%
\begin{figure}
\plottwo{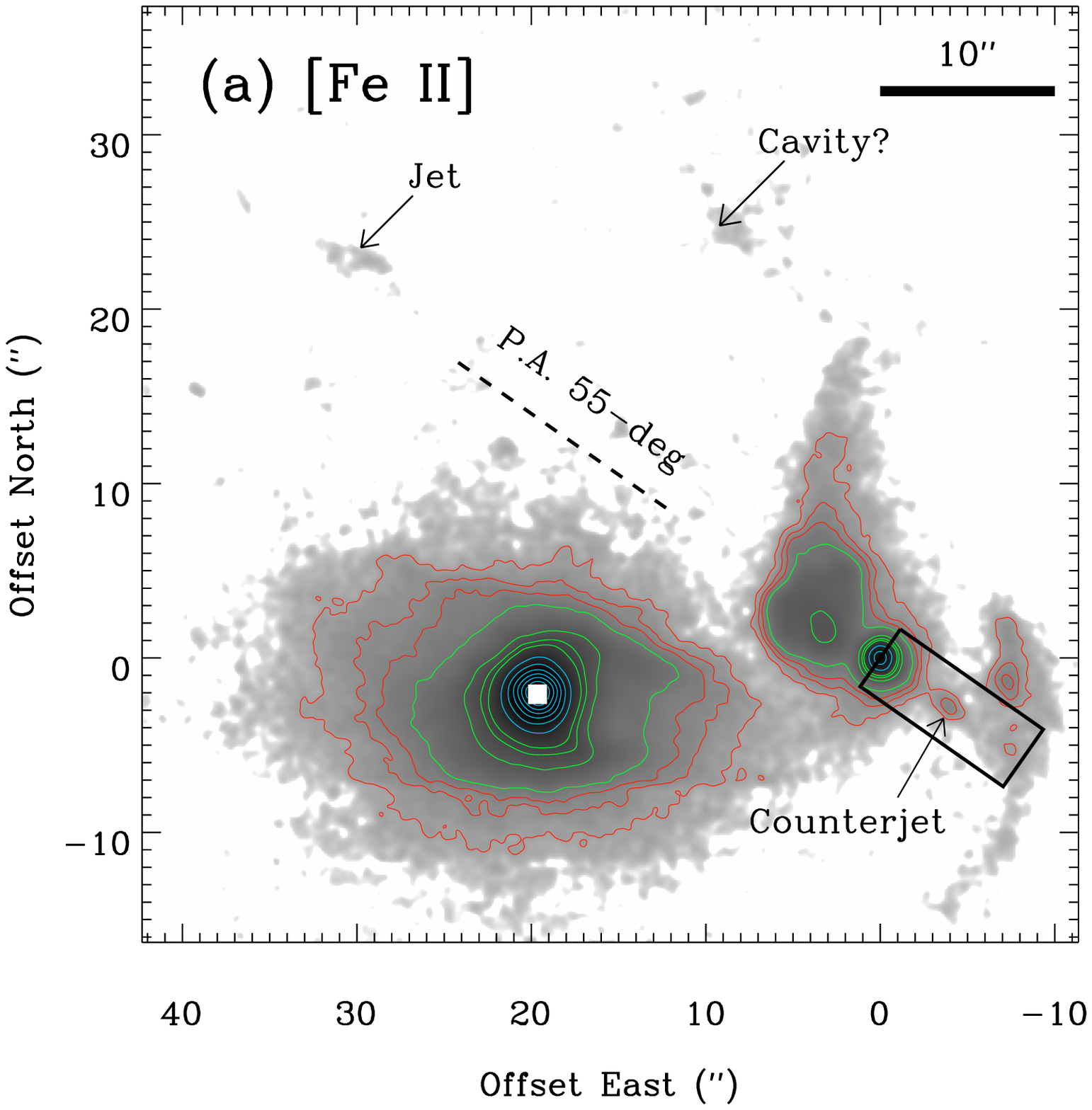}{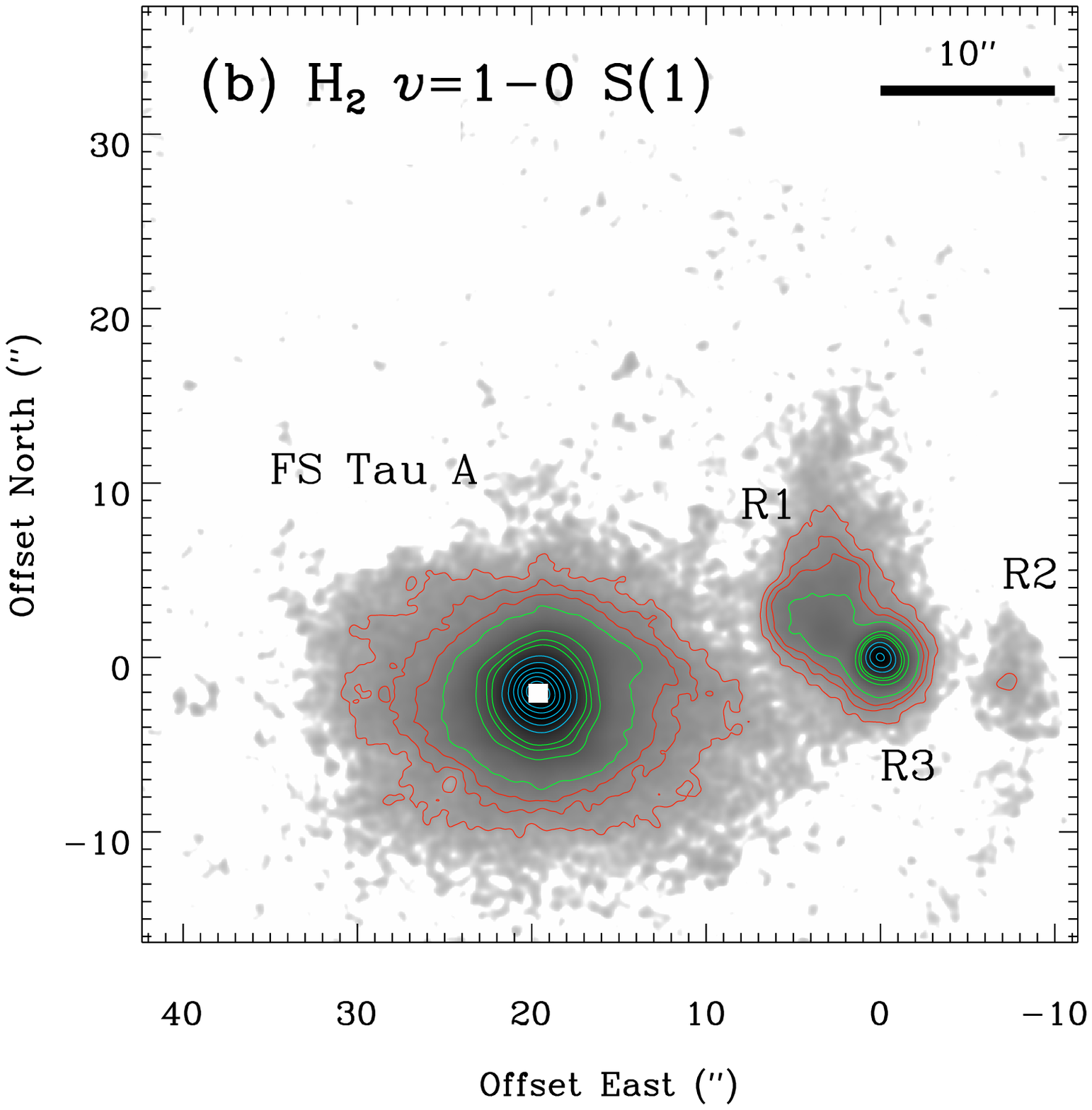} \\
\epsscale{0.8}
\plotone{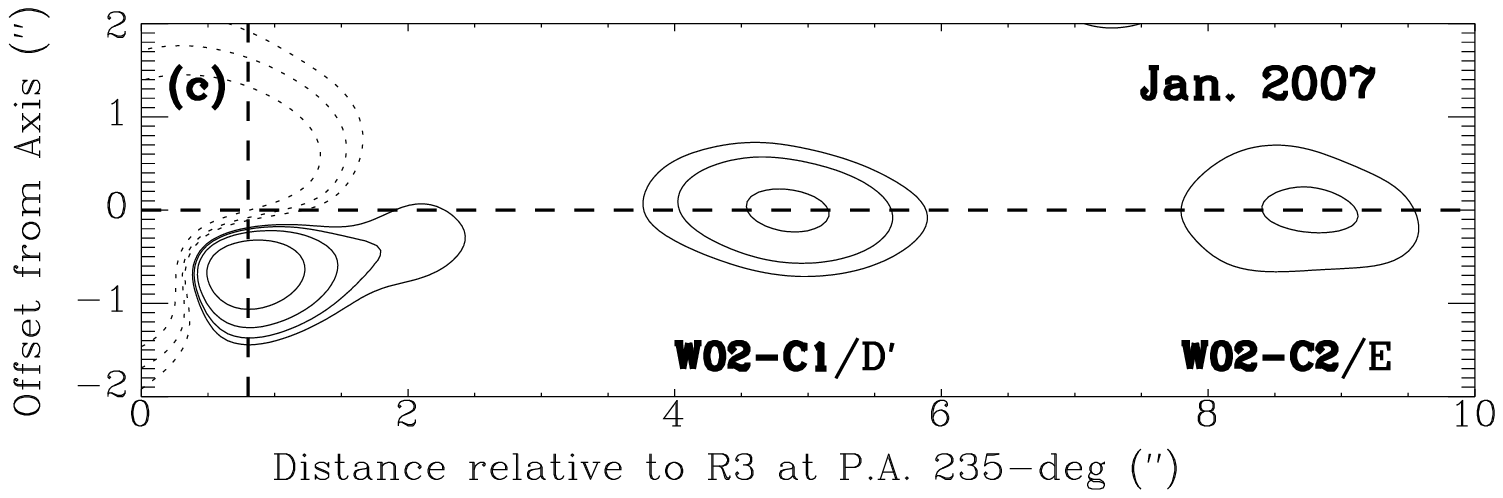}
\epsscale{1.0}
\caption{The Subaru/IRCS images of FS Tau A/B taken with 
         (a) [Fe {\sc ii}] and (b) H$_2$ $(1-0)$ filters. In the H$_2$ image, 
         FS Tau A and the reflection nebulae associated with FS Tau B (R1, R2, and R3) are labeled.
         Each panel has a field of view $53\farcs7 \times 53\farcs7$ and (0,0) is set at the peak
         position of R3. The figures were convolved with a Gaussian of 10-pixel (0\farcs52) FWHM 
         for presentation purpose. The intensities are displayed with logarithmic scale. 
         The central 1\arcsec\ of FS Tau A peak position has been blocked due to saturation.
         The root-mean-square noise ($\sigma$) was determined on each image and contours are 
         superimposed as 3, 6, 9 $\sigma$ (red), 15, 45, 75, 105 $\sigma$ (green), and 
         $2^{i}\times250$ $\sigma$ where $i=0, ..., 10$ (cyan). In the [Fe {\sc ii}] image, the 
         position angle of HH 157 ($\sim 55\degr$) is indicated. (c) The [Fe {\sc ii}] counterjet 
         after continuum subtraction, which is the zoomed-in view of the black rectangle shown in (a).}
\label{IRCSFigs}
\end{figure}

 Fig. \ref{IRCSFigs} shows the Subaru/IRCS images in the
 (a) [Fe {\sc ii}] and (b) H$_2$ narrow-band filters. The images were 
 smoothed with a Gaussian of 0\farcs5 FWHM in order to show the diffuse structures at higher 
 signal-to-noise without degrading the angular resolution. The bright sources are labeled 
 in Fig. \ref{IRCSFigs}(b), including FS Tau A, and reflection nebulae associated with 
 FS Tau B, named as R1 (the NE nebula), R2 (the SW nebula), and R3 (the knot-like nebula), after 
 Eisl\"offel \& Mundt (1998). The optically-invisible FS Tau B lies $\sim 0\farcs8$ 
 southwest to R3 in {\it HST} image (Eisl\"offel \& Mundt 1998; Krist et al. 1998).
 In the IRCS images, it lies within the brightest region of the R3 nebula.

 We detected [Fe {\sc ii}] emission associated with the counterjet of FS Tau B.
 By comparing IRCS images of different filters, we found that an elongated structure appears only 
 in the [Fe {\sc ii}] frame [Fig. \ref{IRCSFigs}(a)] $\sim 5\arcsec$ southwest to R3, at 
 a position angle of 235\degr.
 This direct image of the [Fe {\sc ii}] counterjet confirms the 
 faint detection in the {\it HST}/NICMOS broad-band imaging suggested in Padgett et al. (1999).
 The emission is unresolved in the transverse direction. No [Fe {\sc ii}] 
 emission was detected beyond R2. On the other hand, following the position angle of 55\degr, 
 only a faint structure appears at $\sim 30\arcsec$ northwest from FS Tau B, 
 at the position of HH 157. The peak brightness of this knot is $\sim 5$ times fainter
 than that of the counterjet knot, and its total flux is $\sim 7$ times lower than that of the 
 counterjet.
 
 We subtracted the scaled $H$-band image from the [Fe {\sc ii}] image 
 to obtain the pure line emission
 map for the counterjet of FS Tau B, as shown in Fig. \ref{IRCSFigs}(c).
 The brighter knot, which peaks 5\arcsec\ 
 from R3, is the most obvious structure of the counterjet shown in Fig. \ref{IRCSFigs}(a).
 Continuum subtraction reveals another more diffuse knot at $\sim 8\farcs5$, close to 
 the projected position of R2. 

 We compare our [Fe {\sc ii}] image taken in January 2007 with the [S {\sc ii}] images taken
 in November 1990 (Eisl\"offel \& Mundt 1998) and in August 1997 (Woitas et al. 2002a) 
 to identify the knots. Eisl\"offel \& Mundt (1998) identified three knots (D, E, and F) 
 in the counterjet while Woitas et al. (2002a) detected only two knots, which we label as 
 W02-C1 and W02-C2, respectively.
 We apply proper motions of the three knots (Eisl\"offel \& Mundt 1998) to
 the two images in the later epochs. The brighter [Fe {\sc ii}] knot can be identified 
 with W02-C1 and with knot D if the mean proper motion is
 $\sim 0\farcs235$\,yr$^{-1}$, comparable to the average proper motion in the counterjet.
 The much smaller proper motion Eisl\"offel \& Mundt (1998) reported for knot D could 
 have been influenced by the proximity to the driving source.
 If the fainter [Fe {\sc ii}] knot has the proper motion of knot E,
 then it can be identified with W02-C2.
 Knot F was not detected in Woitas et al. (2002a) 
 and is outside our field of view. We refer to the two [Fe {\sc ii}] knots in the counterjet 
 as D$^\prime$ (for its newly derived proper motion) and E, respectively. In the fainter blueshifted
 jet, only knot A can be identified if an average proper motion of $\sim 0\farcs42$\,yr$^{-1}$
 is adopted (Eisl\"offel \& Mundt 1998). Knot B is not detected in either the [S {\sc ii}] image
 (Woitas et al. 2002a) or in our [Fe {\sc ii}] image.
 The results of knot identification are summarized in Table \ref{KnotID}.
 
%%% Table 1 %%%
\begin{deluxetable}{ccccccccccc}
\tablecolumns{11} 
\tablewidth{0pc} 
\tablecaption{Knot Identification of FS Tau B Jet$^\dag$}
\tablehead{ 
\multicolumn{4}{c}{Eisl\"offel \& Mundt (1998; [S {\sc ii}])} &   \colhead{}   & 
\multicolumn{3}{c}{Woitas et al. (2002a; [S {\sc ii}])} & \colhead{} & 
\multicolumn{2}{c}{This work ([Fe {\sc ii}])} \\ 
\cline{1-4} \cline{6-8} \cline{10-11} \\ 
\colhead{Knot} & \colhead{proper motion$^\S$} & \colhead{1990} & \colhead{2007} & \colhead{} &
\colhead{Knot} & \colhead{1997} & \colhead{2007} & \colhead{} & 
\colhead{Knot} & \colhead{2007} \\
\colhead{} & \colhead{(km\,s$^{-1}$)} & \colhead{(\arcsec)} & \colhead{(\arcsec)} & \colhead{} &
\colhead{} & \colhead{(\arcsec)} & \colhead{(\arcsec)} & \colhead{} &
\colhead{} & \colhead{(\arcsec)}}
\startdata 
\cutinhead{Blueshifted Jet}
A      & $280\pm29$ & 33.3 & 40.0 & & A       & 35.2    & 39.5    & & A          & 38.2    \\
B      & $389\pm57$ & 18.3 & 27.7 & & \nodata & \nodata & \nodata & & \nodata    & \nodata \\
\cutinhead{Redshifted Counterjet}
D      &  $17\pm5$  &  1.9 &  2.3 & &         & \\
$\ast$ & $157\pm10$ &      &  5.9 & & W02-C1  & 2.8     & 5.2     & & D$^\prime$ & 5.0     \\
E      & $158\pm7$  &  5.0 &  8.8 & & W02-C2  & 6.6     & 8.9     & & E          & 8.5     \\
F      & $159\pm9$  &  8.2 & 12.1 & & \nodata & \nodata & \nodata & & \nodata    & \nodata \\
\enddata 
\\
$^\dag$Distance are relative to R3 continuum; assuming FS Tau B being 
       $0\farcs8$ southwest (Eisl\"offel \& Mundt 1998).
\\
$^\S$Assuming a distance to FS Tau B of 140 pc.
\\
$^\ast$Epoch 2007 positions for knot D assuming a mean proper motion 
       of 0\farcs235\,yr$^{-1}$.
\label{KnotID} 
\end{deluxetable} 

\subsection{Kinematics of FS Tau B Jet in [Fe {\sc ii}] Line}

%%% Figure 2 %%%
\begin{figure}
\plotone{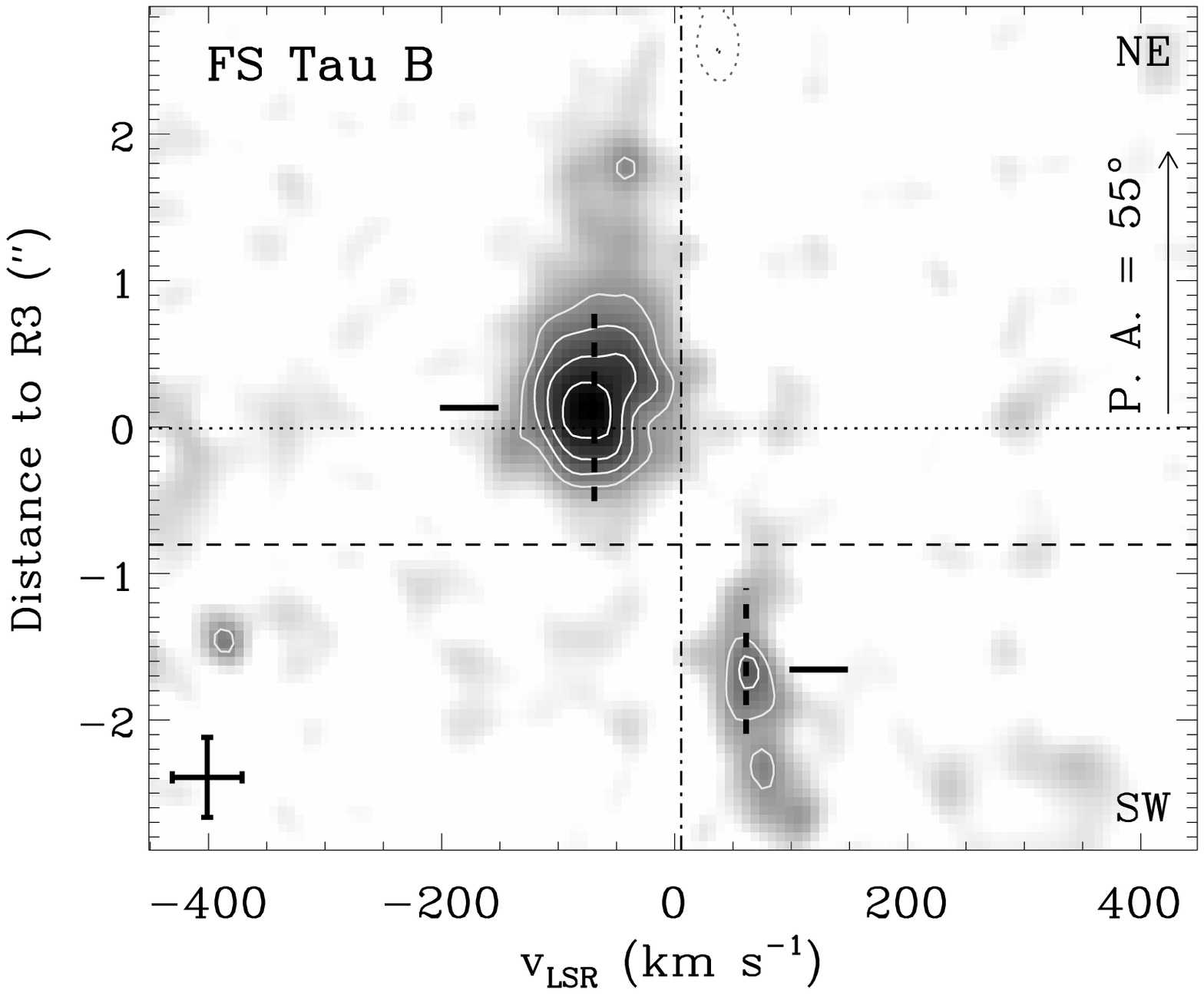}
\caption{Position-velocity diagram of the FS Tau B jet along the jet axis in the [Fe {\sc ii}] 
         1.644 \micron\ line. The position angle of the slit was 55\degr. 
         The data were smoothed by a two-dimensional Gaussian with $\sigma_{\rm G} = $ 12.8 km\,s$^{-1}$
         and 0\farcs115 in the spectral and spatial direction, respectively. 
         The ordinate represents the position relative to the fitted position of the R3 continuum
         peak (indicated by the dotted horizontal line) along the position angle of the slit. 
         The thin dashed line indicates the approximate position of FS Tau B
         (at $-0\farcs8$; Eisl\"offel \& Mundt 1998). 
         The abscissa shows the velocity relative to the local standard of rest.
         The adopted systemic velocity of FS Tau B is $V_{\rm LSR}\sim +6.9$ km\,s$^{-1}$ 
         (Ungerechts \& Thaddeus 1987), labeled by the dash-dotted vertical line. 
         The bars in the lower left corner indicate the velocity resolution and seeing
         (at a FWHM of 60 km\,s$^{-1}$ and 0\farcs55, respectively).
         Positive contours are shown in solid lines and negative contours are shown in dotted lines. 
         The contour levels shown here are $-4$, $-2$, $3$, $4$, $5$, and $6$ times the rms noise 
         ($\sigma \sim 0.003 \times 10^{-18}$ W m$^{-2}$ \AA$^{-1}$). 
         Negative values between $v_{\rm LSR} = 0$ and $+50$ km\,s$^{-1}$ are caused by residuals from
         subtraction of the telluric OH lines. 
         The positions of the brightest knots at the blueshifted and redshifted sides are
         indicated by thick horizontal bars, and velocity centroids from the respectively integration
         of knot in jet and counterjet are denoted by thick vertical dashed lines.
         The blueshifted and redshifted knots have velocities $-74.6$ and $+55.8$ km\,s$^{-1}$
         relative to the systemic velocity, respectively.}
\label{FSTauB_PV}
\end{figure}

 Fig. \ref{FSTauB_PV} shows the position-velocity (p-v) diagram of the FS Tau B jet and counterjet 
 obtained from the continuum-subtracted spectra of [Fe {\sc ii}]. The positions 
 are shown relative to R3, which was determined from the continuum peak. The approximate 
 position of FS Tau B, which is invisible in the $H$-band,
 is at $-0\farcs8$ (Eisl\"offel \& Mundt 1998).
 The emission northeast of FS Tau B is the blueshifted jet and the southwestern
 emission is the redshifted counterjet.

 Within 2\arcsec\ from the driving source, the blueshifted jet appears brighter than the redshifted
 counterjet in [Fe {\sc ii}] line emission. The flux of the brightest knot at the blueshifted 
 and redshifted side is $2.3\times10^{-18}$ and $0.63\times10^{-18}$ W\,m$^{-2}$, respectively.
 This is counter to the larger scale ($\sim 5\arcsec$) behavior as seen in the narrow-band 
 [Fe {\sc ii}] imaging. The brightest blueshifted knot is around 0\farcs15 northeast to the 
 fitted position of R3 nebula. The peak positions of the brightest red and blue knots are 
 0\farcs86 and 0\farcs95 from the source, respectively. The emission becomes faint and noisy 
 as the knots extend $\sim 1\arcsec$ further away from the peak positions.

 The p-v diagram shows that the velocities of the jet and the counterjet of FS Tau B are different. 
 With respect to the systemic velocity, the average velocity for the blueshifted knot is 
 at $-74.6$ km\,s$^{-1}$, and the redshifted one is at $+55.8$ km\,s$^{-1}$, 
 with an uncertainty $\sim 2$ km\,s$^{-1}$.
 This gives a radial velocity ratio of $1.34 \pm 0.05$ between the blueshifted and redshifted sides.
 
 The brightest blueshifted knot and its extension exhibit large line widths, with an average of 
 $\sim 110$ km\,s$^{-1}$.  It decreases down to 80 km\,s$^{-1}$ further away as the knot 
 intensity decreases. On the other hand, the counterjet is not spectrally resolved with a 
 velocity width $\lesssim 50$ km\,s$^{-1}$. The residual of OH airglow subtraction at 
 $v_{\rm LSR} = 0 - 50$ km\,s$^{-1}$ makes it difficult to investigate kinematics in this 
 flow component.

\section{Discussion}

 The combined [Fe {\sc ii}] image and spectra of FS Tau B suggests that it is driving an 
 asymmetric jet. The kinematic properties revealed by the spectra and the intensity difference
 shown in the images indicates that the physical conditions in the two sides of the jet 
 may be different. We combine our new [Fe {\sc ii}] spectroimaging results with previous
 optical data in the literature to provide a more complete picture of the asymmetry.
 
 The [Fe {\sc ii}] spectra provided the first spatially resolved spectra for the FS Tau B jet
 close to the source at a sub-arcsecond scale. Previous optical spectra obtained by 
 Eisl\"offel \& Mundt (1998), at a resolution of $\sim 1\arcsec$, have shown that the 
 velocity ratio of the blueshifted jet to the redshifted counterjet is $\sim 1.24$ and that
 the jet has a larger line width (100 to 200 km\,s$^{-1}$) than the counterjet ($\sim 40$ 
 km\,s$^{-1}$) at the positions close to the source. In our [Fe {\sc ii}] spectra, the 
 speed for both the jet and the counterjet is larger than that found in the [S {\sc ii}] 
 spectra by $\gtrsim 20\%$, resulting in a more pronounced velocity ratio of $1.34\pm0.05$.
 The different velocity ratios derived by lines of different species may result from different
 critical densities that trace different portions of the jet. The [Fe {\sc ii}] line
 has critical density $\sim 10$ times that of [S {\sc ii}] ($2\times10^4$ cm$^{-3}$ compared to
 $2\times10^3$ cm$^{-3}$) and may be able to trace the denser and faster inner part of the jet. 
 Such a difference was also found in the asymmetric RW Aur A jet (e.g., Woitas et al. 2002b; 
 Pyo et al. 2006).
 
The spectrally resolved [Fe {\sc ii}] knot permits a qualitative comparison with wind models. In the $X$-wind models, the wind is driven from the inner edge of the disk, at which the streamlines diverge almost radially with the same flow speed. With a projected jet velocity centroid $v_j$ and inclination $i$, the peak blueshifted velocity will occur at the deprojected speed $v_j/\cos i$. This sets the scale of the line width. The innermost blueshifted knot of the FS Tau B jet has a peak velocity of $-75$ km\,s$^{-1}$ with a FWHM of $\sim 110$ km\,s$^{-1}$. With $i=70\degr$, the deprojected speed is $\sim220$ km\,s$^{-1}$ for the blueshifted jet. The FWHM of $\sim110$ km\,s$^{-1}$ is consistent with the order suggested by the $X$-wind model. Such a large linewidth close to the driving source was also observed in the blueshifted jet from RW Aur A, with a value $\sim 100$ km\,s$^{-1}$ (Pyo et al. 2006). The counterjet would have been expected to have a similar scale of line widths close to the driving source. However, the imperfect subtraction of telluric lines and possible disk occultation (discussed below) would complicate the determination of the line widths at the redshifted side. 

 Flux asymmetry between the jet and counterjet was observed in our [Fe {\sc ii}] image as well as 
 in the optical images obtained with {\it HST}/WFPC2 (Krist et al. 1998; Woitas et al. 2002a).
 For [Fe {\sc ii}] and WFPC2 $R$-band images, the peak brightness ratio between the jet knot A and 
 counterjet knot D$^\prime$ is below 0.3, while that in the H$\alpha$ and [S {\sc ii}] images is 
 $\sim 2$. The apparent discrepancy was attributed to different treatments of pixel scales when
 mosaicking WFPC2 CCDs. With reanalysis with the standard {\tt MultiDrizzle} procedure, the ratio becomes
 $\sim 0.5$ for the narrow-band images. It is puzzling that the [Fe {\sc ii}] flux ratio between 
 the jet and counterjet changes from $1/7$ to $\sim 3$ within 2\arcsec\ from the source. One 
 possibility is the differential extinction between the two sides of the jet. 
 However, extinction values found in the literature (obtained from optical to infrared imaging and 
 spectroscopic observations) are widely scattered, and the observations not spatially resolve the knots (e.g., Krist et al.\ 1998; Padgett et al.\ 1998; Connelley \& Greene 2010). Apparent dimming due to 
 partial disk occultation may explain the flux ratio if one assumes that the two innermost 
 knots were ejected at the same epoch with the 1.34 velocity ratio. An opaque disk with radius 
 $309\pm18$ AU at an inclination of $\sim70\degr$ (Yokogawa et al.\ 2001) is consistent with this scenario.

 The brightness and flux asymmetries observed in both optical and near-infrared wavelengths
 may result from asymmetries in excitation conditions at the two sides of the disk.
 Optical spectra taken by Eisl\"offel \& Mundt (1998) showed that the density-sensitive
 [S {\sc ii}] $\lambda6717/6731$ ratio is lower in the redshifted counterjet, 
 indicating that the redshifted counterjet has an overall higher electron density.
 Woitas et al. (2002a) noted that a flux asymmetry occurs in the H$\alpha$ and
 [S {\sc ii}] images, and that the H$\alpha$/[S {\sc ii}] ratio
 is lower in the redshifted counterjet, suggestive of a lower ionization state. 
 The brighter [Fe {\sc ii}] counterjet also 
 has an overall higher density than the blueshifted jet. 
 Line intensities combined with the line ratios obtained in both the optical and near-infrared 
 wavelengths suggest that FS Tau B is driving a jet that has different density profiles and excitation
 conditions on opposite sides.

 Properties of the FS Tau B system can be explained by asymmetry in the mass-loss driven from the source. 
 The asymmetry in mass-loss rates could cause an asymmetry in velocity in order to balance the linear 
 momentum ejected. That is, a smaller mass-loss would be compensated by a higher jet velocity, 
 and vice versa, because no apparent recoil motion has been reported for such sources. The faster 
 blueshifted jet would have a smaller mass-loss rate which gives rise to fainter [Fe {\sc ii}] 
 emission and a larger [S {\sc ii}] doublet ratio. The faster jet would also carry more kinetic energy 
 that could inject more mechanical heating into the jet leading to a larger value of 
 H$\alpha$/[S {\sc ii}] ratio. On the other hand, material on the cavity wall should experience different 
 amounts of shock heating as shown in the brightnesses of H$\alpha$ along the cavity 
 (Fig. 9 of Eisl\"offel \& Mundt 1998). Ambient material around FS Tau B is likely to be 
 symmetrically distributed as the bipolar cavities are symmetric in morphology (e.g., Eisl\"offel \& Mundt 1998).
 The asymmetry in the jet alone could deposit different amounts of energy onto the walls through the 
 unseen wide-angle wind that the wide velocity profile near the base manifests. The asymmetric optical 
 emission along the cavities may be the evidence of such interaction.

 There are other jet-outflow systems that have an asymmetric jet driving an outflow.
 The Class I source L1551 IRS 5 has a large-scale H$\alpha$ nebula surrounding its blueshifted jet 
 (Davis et al.\ 1995). The redshifted counterjet was not well known due to its high extinction until 
 recent deep [Fe {\sc ii}] imaging (Hayashi \& Pyo 2009). The asymmetry is even more pronounced in that the blueshifted lobe of the molecular outflow carries twice the kinetic energy of the redshifted lobe (Stojimirovi\'c et al. 2006). 
 HL Tau, a Class II source, drives a bipolar [S {\sc ii}] jet with a discernible asymmetry. 
 Velocity of the blueshifted jet is $\sim -200$ km\,s$^{-1}$ while the redshifted counterjet is 
 $\sim +100$ km\,s$^{-1}$ in the stellar rest frame (Mundt et al. 1990). 
A similar asymmetry was also noted in the [Fe {\sc ii}] jet (although with a smaller velocity 
 ratio for yet unknown reasons; Pyo et al. 2006). Transitional objects such as FS Tau B and HL Tau are excellent targets to examine the asymmetry in both the jet and its associated wind-ambient interaction.

\section{Summary}

 We have undertaken Subaru/IRCS observations of the FS Tau B system 
 in the $H$ and $K$ bands with [Fe {\sc ii}] 1.644 \micron\ and H$_2$ 2.122 \micron\ lines.
 We have for the first time detected the jet and counterjet in [Fe {\sc ii}] images, 
 and spatially resolved the kinematics of the jet and counterjet
 within 2\arcsec\ of the driving source.
 The [Fe {\sc ii}] jet and counterjet clearly exhibit asymmetric properties. 
 At distances $> 5\arcsec$ from the driving source, the total flux of the counterjet is 7 times
 that of the jet, while within 2\arcsec the blueshifted jet is brighter by a 
 factor of 3. The innermost blueshifted knot is 34\% faster and shows
 a larger velocity width than the innermost redshifted knot. 

 We compare our [Fe {\sc ii}] imaging and spectra with optical results of archival {\it HST}/WFPC2 
 imaging and long-slit spectra in the literature. The [Fe {\sc ii}] images show a stronger 
 brightness asymmetry than do the optical narrow-band and broad-band images. 
 The asymmetry cannot be due solely to differential extinction. The fainter
 and faster [Fe {\sc ii}] jet shows the larger [S {\sc ii}] doublet ratio and H$\alpha$/[S {\sc ii}] 
 ratio, indicating that it has lower density and higher excitation. This suggests the FS Tau B
 jet and counterjet are asymmetric. Interpretation of the observed asymmetries of
 the innermost knots is complicated by possible differential extinction and disk obscuration.

 The FS Tau B system may be explained by an asymmetric bipolar outflow driven with different 
 mass-loss rates. The side with a lower mass-loss rate will drive a jet with a higher speed
 to maintain momentum balance. Thus the faster and less dense jet can produce fainter
 [Fe {\sc ii}] emission but a larger H$\alpha$/[S {\sc ii}] ratio since it carries more kinetic
 energy. The difference in brightness of H$\alpha$ knots seen on two sides of the cavity can also be 
 explained as different amounts of energy impinging onto the wall, giving rise to higher
 ionization on the faster side. This naturally produces the observational results without any assumptions about the properties of the ambient medium. This confluence of detailed observations and models makes FS Tau B a promising laboratory for testing models of the origins of protostellar outflows.

{\it Facilities:} \facility{Subaru (IRCS)}

\acknowledgments

The authors thank the anonymous referee whose comments significantly improved presentation and clarity of the manuscript. This work was supported by funds from the Academia Sinica Institute of Astronomy and Astrophysics, and National Science Council (NSC) of Taiwan by grants NSC97-2112-M-001-018-MY3 and NSC96-2752-M-001-001-PAE.

\end{document}